\documentclass{article}

\usepackage{arxiv}

\usepackage[utf8]{inputenc} 
\usepackage[T1]{fontenc}    
\usepackage{hyperref}       
\usepackage{url}            
\usepackage{booktabs}       
\usepackage{amsfonts}       
\usepackage{nicefrac}       
\usepackage{microtype}      
\usepackage{lipsum}
\usepackage{graphicx}
\usepackage{multirow} 
\usepackage{graphicx} 

\usepackage{geometry}
\usepackage[ruled,vlined]{algorithm2e} 
\usepackage{amsmath}
\graphicspath{ {./images/} }

\title{High-Precision Medical Speech Recognition Through Synthetic Data and Semantic Correction: United-MedASR}

\author{
    \begin{tabular}{ccc}
        \textbf{Sourav Banerjee}\thanks{To whom correspondence should be addressed: E-mail: sb@unitedwecare.com} & \textbf{Ayushi Agarwal} & \textbf{Promila Ghosh} \\
        \normalfont DataLabs, United We Care & \normalfont DataLabs, United We Care & \normalfont DataLabs, United We Care \\
    \end{tabular}
}
\date{}

\begin{document}
\maketitle
\begin{abstract}
Automatic Speech Recognition (ASR) systems in the clinical domain face significant challenges,
notably the need to recognise specialised medical vocabulary accurately and meet stringent precision
requirements. We introduce United-MedASR, a novel architecture that addresses these challenges by
integrating synthetic data generation, precision ASR fine-tuning, and advanced semantic enhancement
techniques. United-MedASR constructs a specialised medical vocabulary by synthesising data from
authoritative sources such as ICD-10 (International Classification of Diseases, 10th Revision), MIMS
(Monthly Index of Medical Specialties), and FDA databases. This enriched vocabulary helps fine-tune the Whisper ASR model to better cater to clinical needs. To enhance processing speed,
we incorporate Faster Whisper, ensuring streamlined and high-speed ASR performance. Addition-
ally, we employ a customised BART-based semantic enhancer to handle intricate medical terminology,
thereby increasing accuracy efficiently. Our layered approach establishes new benchmarks in ASR
performance, achieving a Word Error Rate (WER) of 0.985\% on LibriSpeech test-clean, 0.26\% on
Europarl-ASR EN Guest-test, and demonstrating robust performance on Tedlium (0.29\% WER) and
FLEURS (0.336\% WER). Furthermore, we present an adaptable architecture that can be replicated across different domains, making it a versatile solution for domain-specific ASR systems.
 
\end{abstract}


\section{Introduction}

While Automatic Speech Recognition (ASR) excels in general transcription tasks \cite{1}, domain adaptation presents unexplored challenges in speech processing. State-of-the-art ASR frameworks enable real-time transcription across multiple sectors. However, current encoder-decoder architectures with attention mechanisms show performance limitations in specialized domains \cite{1}. These limitations manifest when processing domain-specific terminology, resulting in increased error rates across healthcare, legal, and scientific fields \cite{2}. Domain-specific ASR systems require specialized vocabulary processing, contextual understanding, and high accuracy transcription capabilities that exceed current architectural designs \cite{3}\cite{4}. Current leading models, including OpenAI's Whisper, Google's Speech-to-Text, and Microsoft Azure Speech \cite{5}-\cite{8}, while proficient in everyday scenarios, often falter when confronting complex medical terminology. The accurate transcription of drug names, anatomical terms, and clinical procedures requires a level of specialized understanding that general-purpose ASR systems typically lack, potentially leading to critical errors in medical settings.
These challenges are not unique to healthcare but resonate across numerous domains requiring specialized vocabulary training. Legal proceedings, technical documentation, scientific research, and many other fields face similar needs for domain-specific ASR systems capable of accurately handling specialized terminology and context-specific language patterns. The common thread across these domains is the requirement for highly accurate, context-aware speech recognition that can reliably process field-specific terminology and conventions.
The creation of speech recognition systems tailored to domains encounters significant challenges that mainly revolve around obtaining and organizing data efficiently. This procedure can be costly as it demands investments, in recording devices and skilled annotators of precisely tagging the data. Additionally it is a time consuming process, in sectors where expert validation is crucial, for upholding data accuracy. Moreover gathering and labeling human audio data are frequently hindered by privacy laws and confidentiality stipulations render methods of speech recognition system development unsustainable. The lack of top notch training data tailored to domains hinders the progress of ASRs, in areas significantly. 

To tackle these obstacles head-on we present United-MedASR. A design that merges data creation precise ASR adjustments and cutting-edge semantic improvement features. Our system stands out by utilizing references such, as ICD-10, MIMS and FDA repositories to form a thorough professional glossary. This structure includes Faster Whisper, for enhanced efficiency. Integrates a specialized BART-derived semantic enhancer crafted especially for medical jargon. This holistic method enables us to address the challenges posed by data availability while upholding the identification of specialized terms, with a high level of accuracy.

United-MedASR achieves sub-1\% error rates across domains and datasets : 0.985\% (LibriSpeech), 0.26\% (Europarl-ASR), 0.29\% (Tedlium), and 0.336\% (FLEURS). Word Error Rate (WER) measures speech recognition accuracy by comparing an actual reference text to the predicted text. 
This study dives into our method, for creating data sets and improving model efficiency in speech recognition technology within the medical field while ensuring flexibility for different sectors of expertise. We showcase the success of United-MedASR and how it tackles the obstacles to medical speech recognition while being versatile across industries. This research does not push the boundaries of medical speech recognition. It also lays the foundation for constructing resilient and specialized speech recognition systems that are applicable, in various niche areas.

\section{Related Works}

\subsection{Evolution of ASR Technology}
The development of Automatic Speech Recognition (ASR) began with systems using Hidden Markov Models (HMMs) combined with Gaussian Mixture Models (GMMs). These models were foundational in early ASR but struggled with accuracy in noisy environments and large vocabularies. The introduction of Deep Neural Networks (DNNs) improved acoustic modelling but required manual feature engineering. Later, Recurrent Neural Networks (RNNs) and Long Short-Term Memory (LSTM) models helped capture sequential dependencies in speech, though they faced scalability issues \cite{13}. Early ASR systems operated offline, processing entire audio inputs at once, which made them unsuitable for real-time use. They also relied heavily on predefined language models, limiting their effectiveness in noisy environments and their ability to manage long-range dependencies in speech \cite{14}.

\subsection{Transition to Transformer-Based ASR Models}
Shifting to transformer-based ASR models, driven by attention mechanisms, revolutionised speech processing. Whisper, trained on 680,000 hours of multilingual audio, generalises effectively across languages, nearing human-level accuracy without fine-tuning \cite{14}. While Fadam isn't transformer-based, it optimises models like Adam by leveraging second-order information to improve training efficiency, which benefits transformer models \cite{15}. Models such as Wav2Vec 2.0, Conformer, and Hubert have further enhanced ASR, especially in low-resource and noisy environments \cite{16}, \cite{17}, \cite{18}. However, challenges in achieving naturalness, particularly with synthetic data, remain key obstacles for future improvements.
Mamba, a state-space model, offers a compelling alternative to Transformers, especially in real-time ASR. With linear complexity, Mamba outperforms the quadratic complexity of Transformers in handling streaming data, making it more computationally efficient for ASR. In \cite{19}, Mamba's implementation for streaming ASR demonstrates competitive performance in accuracy and latency, utilising a lookahead mechanism to exploit future information. The introduction of Unimodal Aggregation (UMA) further enhances token representation, minimising recognition delays. Tests on Mandarin datasets confirm Mamba’s efficiency, suggesting it is a strong contender against Transformer-based models for ASR tasks.

\subsection{Evolution of Synthetic Data in ASR systems}
Synthetic data has advanced clinical ASR applications, offering scalable, privacy-preserving datasets crucial for medical transcription and patient monitoring. This is particularly important in addressing data scarcity for rare medical conditions, where real-world data is hard to obtain. By generating synthetic speech data, ASR models are trained on diverse scenarios, improving their ability to recognise medical terminologies while maintaining the confidentiality of sensitive health records \cite{11}, \cite{17}, \cite{18}.
Generative models like Diffusion models and GANs have further enhanced ASR by simulating patient speech data, allowing systems to operate in data-scarce environments. These models enable ASR to better recognise speech patterns linked to rare medical conditions, improving robustness in specialised medical contexts \cite{19}, \cite{20}, \cite{21}.
However, synthetic data presents limitations. It often lacks the complexity and variability of real-world audio, which can degrade ASR accuracy in noisy or unpredictable settings where speech nuances are critical \cite{22}. Ethical concerns, such as the risk of de-anonymisation, also persist, especially in medical applications where patient privacy is paramount \cite{23}\cite{24}.
Despite these challenges, synthetic data remains vital in advancing clinical ASR. Enhancing the realism and variability of synthetic datasets is essential for future success, enabling ASR systems to overcome current limitations in medical and specialised domains.

\section{Material and Method}
\subsection{Overview}
\begin{figure}[h]
    \centering
    \includegraphics[width=0.8\textwidth, height=0.5\textheight]{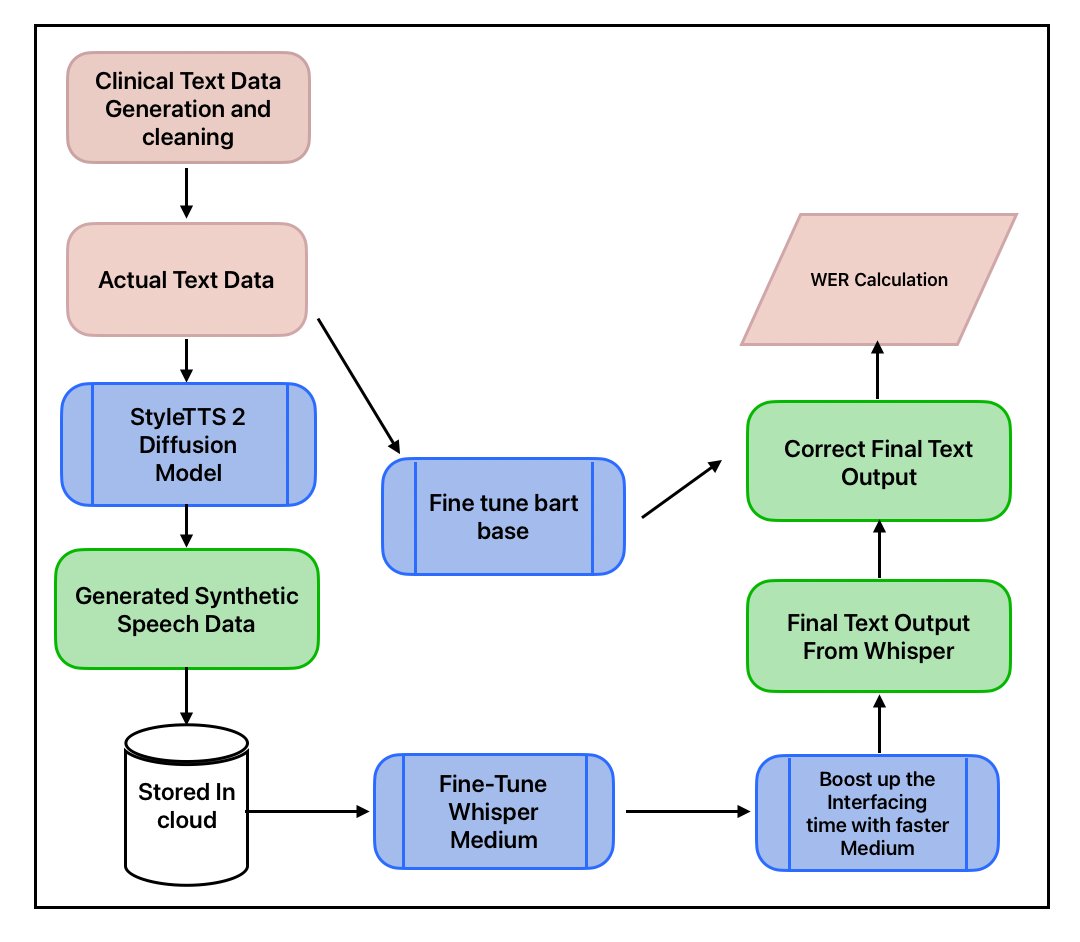}
    \caption{End-to-End Workflow of United-MedASR ASR System Development}
    \label{fig1}
\end{figure}

The development of the United-MedASR system for medical data follows a meticulously designed methodology to ensure accuracy and efficiency. It begins by gathering reliable medical data from trusted sources such as ICD-10, MIMS, and the FDA \cite{25}-\cite{27}. A web scraping tool extracts relevant text and audio, which is then processed by a GPT model to generate realistic medical sentences. These sentences are converted into high-quality synthetic speech using StyleTTS \cite{22}.
The synthetic speech is input into a fine-tuned Speech-to-Text system based on the Whisper model, optimised to recognise complex medical terms. To enhance performance, the model is upgraded to Faster-Whisper, improving speed and accuracy. A semantic enhancement model is integrated to correct transcription errors, particularly for complex medical vocabulary, ensuring precision in the final output. Each step is designed to create a robust, accurate, and efficient ASR system, as depicted in Figure \ref{fig1}.

\subsection{Data and Baseline}

The project begins with the crucial step of data collection, where medical data is sourced from reputable and authoritative platforms such as ICD-10, MIMS-India, and the FDA websites. This step-by-step process makes the ASR system highly effective for use in medical settings as Figure \ref{fig1}. These sources are selected for their reliability and relevance to the medical field. The data collection process is facilitated by using Beautiful Soup, a Python library that enables web 
scraping. This tool is employed to extract relevant medical terminologies, descriptions, and additional pertinent information from the web pages. The extracted text data is subsequently processed to generate contextually accurate sentences using the Generative Pre-trained Transformer (GPT) model [9]. This step ensures that the content generated is not only relevant but also tailored to the specific requirements of the medical domain. Each generated sentence is systematically assigned a unique identifier (ID) to facilitate organisation, tracking, and retrieval during subsequent stages of the project. 
\\
\begin{algorithm}[H]
\caption{Generate Synthetic Text Function}
\SetKwFunction{FGenerateSyntheticText}{GenerateSyntheticText}
\KwIn{$D_{scraped}$: Collected data}
\KwOut{$T_{syn}$: Generated synthetic text}
$M_{GPT} \gets \textsc{LoadGPTModel}()$\;
$T_{syn} \gets \emptyset$\;
\For{$t \in D_{scraped}$}{
    $c \gets \textsc{GenerateContext}(t, M_{GPT})$\;
    $s \gets \textsc{GenerateSentence}(t, c, M_{GPT})$\;
    $id \gets \textsc{GenerateUUID}()$\;
    $T_{syn} \gets T_{syn} \cup \{(id, s)\}$\;
}
\Return $T_{syn}$\;
\end{algorithm}
\subsection{Audio Data Generation Techniques}
Following the generation of sentences, the next phase involves converting these text sentences into synthetic speech. This process is applied to a substantial dataset comprising 60,000 sentences from F-Series medical data and 335,000 sentences from MIMS and FDA data, resulting in a total of 395,000 unique sentences. These sentences are systematically saved with unique names to ensure that they can be efficiently catalogued and retrieved when necessary. The conversion from text to speech is accomplished using StyleTTS 2, a state-of-the-art diffusion Text-to-Speech (TTS) model. StyleTTS 2 utilises advanced techniques such as style diffusion and adversarial training in conjunction with large speech language models (SLMs) to produce high-quality synthetic speech \cite{22}. The audio generation process is meticulous. 
The alpha parameter controls the blending of the reference style with the predicted style for the first half of the style vector. A higher alpha value indicates greater reliance on the reference style, which can influence the tonal and prosodic characteristics of the synthesised speech. Beta (0.7) is similar to alpha, beta affects the second half of the style vector, determining the degree to which the predicted style influences the final style representation in the synthesised speech. Diffusion Steps (6)  specifies the number of diffusion steps used to generate or refine style embeddings. More steps generally result in more refined and accurate style representations, contributing to the overall quality of the synthesised speech. An Embedding Scale (1), scaling factor is applied to the embeddings during the diffusion process. It influences the magnitude of the embeddings, which can have an impact on the final quality of the synthesised speech. These parameters are carefully applied during the inference process of StyleTTS to generate approximately 790,000 TTS files, which collectively amount to around 5,486 hours of labelled audio, featuring both male and female voices, and store them at Google Cloud Storage (GCS) \cite{28}. Each of these audio files is standardised to a length of 30 seconds to comply with the requirements of the Whisper model, which will be used in subsequent stages for Speech-to-Text conversion.
The dataset, named United-Syn-Med, is now publicly accessible on HuggingFace, providing a valuable resource for researchers and developers in the medical and AI communities who seek high-quality, labelled medical speech data for model training and analysis \cite{49}.

\begin{algorithm}[H]
\caption{Generate Synthetic Audio Function}
\SetKwFunction{FGenerateSyntheticAudio}{GenerateSyntheticAudio}
\KwIn{$T_{syn}$: Generated synthetic text}
\KwOut{$A_{syn}$: Generated synthetic audio}
$M_{TTS} \gets \textsc{LoadStyleTTS2Model}()$\;
$A_{syn} \gets \emptyset$\;
\For{$(id, s) \in T_{syn}$}{
    $a \gets M_{TTS}(s, \alpha=0.3, \beta=0.7, \text{diffusion\_steps}=6)$\;
    $a_{std} \gets \textsc{StandardiseAudio}(a, \text{duration}=30)$\;
    $A_{syn} \gets A_{syn} \cup \{(id, a_{std})\}$\;
}
\Return $A_{syn}$\;
\end{algorithm}

\subsection{Data Preprocessing for Whisper Model Training}
In this phase, the synthetic speech files generated earlier, along with the mapped IDs from the previously generated sentences, are formatted as a dataset compatible with Transformers models. The data is stored on disk to ensure accessibility, and for consistency and reproducibility, it is split into training and testing sets in an 80-20 ratio, with the random state parameter set to 42 \cite{29}.

\begin{algorithm}[H]
\caption{Preprocess Data Function}
\SetKwFunction{FPreprocessData}{PreprocessData}
\KwIn{$A_{syn}$: Generated synthetic audio, $T_{syn}$: Generated synthetic text}
\KwOut{$D_{train}, D_{test}$: Processed training and testing data}
$D_{processed} \gets \textsc{FormatForTransformers}(A_{syn}, T_{syn})$\;
$D_{train}, D_{test} \gets \textsc{TrainTestSplit}(D_{processed}, \text{ratio}=0.8)$\;
\Return $D_{train}, D_{test}$\;
\end{algorithm}

\subsection{Fine-Tuning Process of the Whisper}

\begin{figure}[h]
    \centering
    \includegraphics[width=0.8\textwidth, height=0.3\textheight]{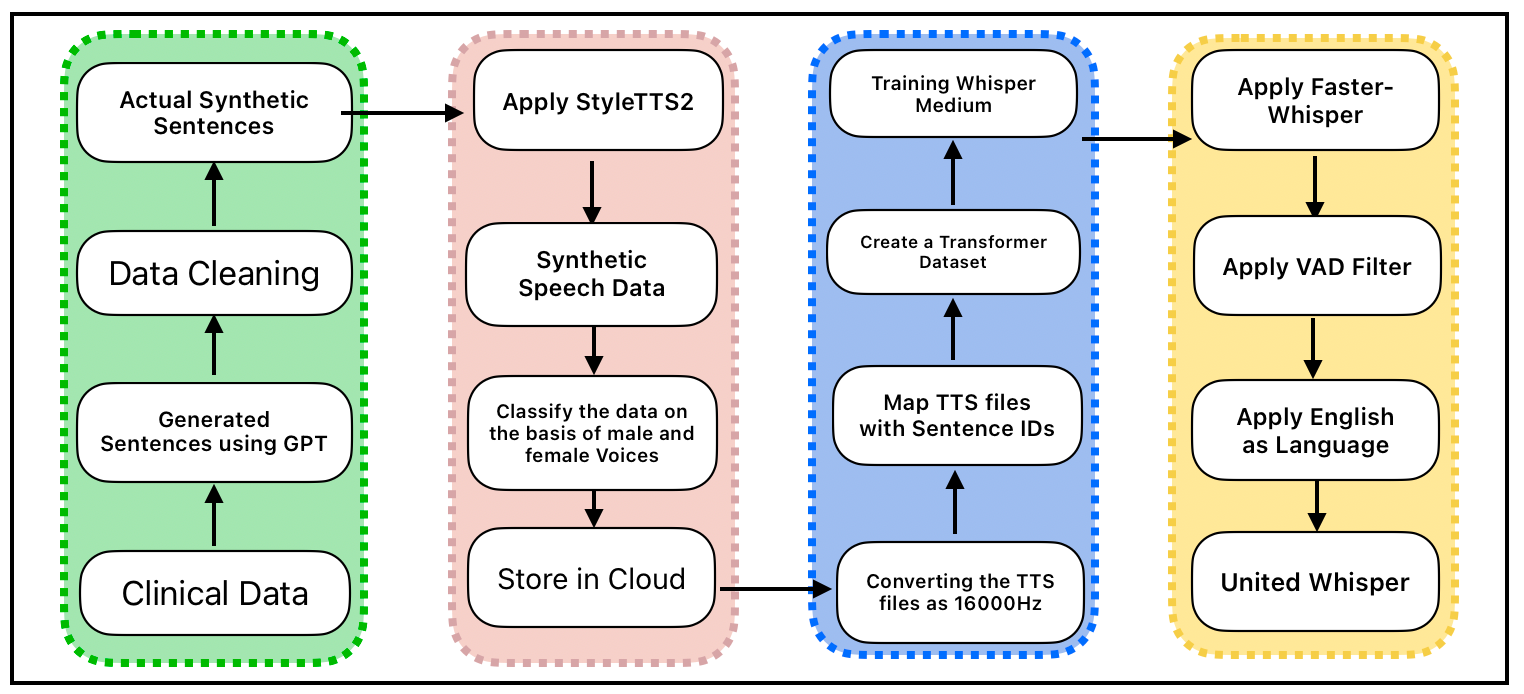}
    \caption{Synthetic Data Pipeline and United-MedASR Training Process
}
    \label{fig2}
\end{figure}

To fine-tune the Whisper medium model \cite{15} for improved accuracy and adaptability in medical speech recognition, the process begins with data preparation. A data preparation function is defined to clean and prepare the medical text data, ensuring that the data used for training is of high quality and relevant to the task. The function processes each batch of audio data by first loading and resampling the audio to 16kHz. It then computes input features, specifically log-Mel spectrograms, from the audio array for use in model training. Additionally, the target text is tokenised into label IDs to create a corresponding output for the model. Once these steps are completed, the function returns the batch with the extracted features and encoded labels, preparing it for further use in machine-learning tasks.  Both the training and testing datasets are loaded from disk, alongside pre-trained components such as the feature extractor, tokeniser, processor, and model from the Hugging Face Transformers library \cite{30}.  The data collator processes our pre-prepared data and converts it into PyTorch tensors, ready for the model. A custom data collator class is implemented to manage padding during batch processing, making sure that all sentences in a batch are of uniform length for efficient processing. Figure \ref{fig2} depicts the workflow for generating synthetic speech data and using it to train the Whisper medium model with Faster-Whisper, optimizing real-time processing performance.

The training process is managed using the Trainer class from the Transformers library, which handles interactions between the model, prepared dataset, tokeniser, and a custom data collator \cite{34}.  The Word Error Rate (WER) metric is used to evaluate model performance during training and testing.
The Word Error Rate (WER) is calculated using the following equation:
\begin{equation}
\text{WER} = \frac{S + D + I}{N}
\end{equation}

\noindent Where:

\begin{itemize}
    \item $S$ = Number of substitutions (words that were incorrectly transcribed)
    \item $D$ = Number of deletions (words that were missed or omitted)
    \item $I$ = Number of insertions (extra words that were incorrectly added)
    \item $N$ = Total number of words in the reference text (the original, correct text)
\end{itemize}

\noindent The WER is typically expressed as a percentage, indicating the proportion of words that were incorrectly transcribed by the model.

\begin{algorithm}[H]
\caption{Fine-Tune Whisper Model Function}
\SetKwFunction{FFineTuneWhisperModel}{FineTuneWhisperModel}
\KwIn{$D_{train}, D_{test}$: Training and testing data}
\KwOut{$M_{FT}, \text{WER}$: Fine-tuned model and WER}
$M_W \gets \textsc{LoadWhisperMedium}()$\;
$C \gets \textsc{CustomDataCollator}()$\;
$\text{TrainingArgs} \gets \{\text{batch\_size}: 16, \text{lr}: 1e-5, \text{epochs}: 10\}$\;
$M_{FT} \gets \textsc{Train}(M_W, D_{train}, C, \text{TrainingArgs})$\;
$\text{WER} \gets \textsc{EvaluateWER}(M_{FT}, D_{test})$\;
\Return $M_{FT}, \text{WER}$\;
\end{algorithm}

The training process is configured with specific parameters: an output directory for saving models and logs, a batch size of 1 per device, a learning rate set at 1\text{e-}5, 500 warm-up steps to stabilise learning and a maximum of 5000 training steps. Logging intervals are set at every 25 steps to monitor progress, with evaluations based on WER occurring every 1000 steps. The model is set to generate predictions up to 225 tokens, and the best-performing model, based on the lowest WER, is saved at each evaluation interval. After fine-tuning, the model's performance is validated using a test dataset, with WER logged as the primary metric to assess the accuracy of transcriptions, particularly in specialised medical domains.

The predictions made by the fine-tuned model are saved for future use, allowing for further analysis and refinement if necessary.
\subsection{Implementation of Faster-Whisper}
To enhance the performance and speed of the ASR system, the fine-tuned whisper model is converted to use Faster-whisper, an optimised version of OpenAI's Whisper model \cite{31}. Faster-whisper leverages a fast inference engine for Transformer models, to accelerate processing. It accelerates speech recognition on CPU and GPU by using advanced optimisations like layer fusion, in-place processing, and batch reordering, achieving faster execution with lower resource demands by CTranslate2. Supporting quantised weights and optimised for multiple CPU architectures, it adapts dynamically at runtime for optimal performance. Parallel, asynchronous processing and dynamic memory management further enhance efficiency, while quantisation reduces model size by up to 4x with minimal accuracy loss. This makes Faster Whisper ideal for high-speed, resource-efficient ASR applications. Additionally, the Silero VAD (Voice Activity Detection) model is integrated to automatically remove non-speech parts of the audio, further improving the model's efficiency in real-time applications. When loading a model by size, such as the Whisper Model, the corresponding fast inference engine for the Transformer model is automatically downloaded from the Hugging Face Hub, ensuring seamless integration and deployment. This conversion process is essential for making the ASR system suitable for real-time applications, where speed and accuracy are paramount.
\begin{algorithm}[H]
\caption{Convert to Faster Whisper Function}
\SetKwFunction{FConvertToFasterWhisper}{ConvertToFasterWhisper}
\KwIn{$M_{FT}$: Fine-tuned Whisper model}
\KwOut{$M_{UW}$: Faster Whisper model}
$M_{FW} \gets \textsc{ConvertToFasterWhisper}(M_{FT})$\;
$M_{VAD} \gets \textsc{LoadSileroVAD}()$\;
$M_{UW} \gets \textsc{IntegrateVAD}(M_{FW}, M_{VAD})$\;
\Return $M_{UW}$\;
\end{algorithm}

\subsection{Fine-Tuning the BART-Base Model for Semantic Enhancement}

The BART-Base model's fine-tuning for semantic enhancement in medical transcription addresses critical challenges in converting spoken medical narratives to structured text \cite{38}. The model was trained on diverse clinical dictations to accurately capture and preserve semantic relationships in medical documentation. For instance, when processing transcripts like "The patient was reviewed today, and we prescribed Amoxicillin 500 mg, 10 tablets, to be taken over a week with three refills," the model must correctly interpret temporal markers, clinical actions, and medication details from natural speech. Its semantic processing capabilities handle complex narratives with multiple medications and varied instructions, as shown in "We also prescribed Ibuprofen 200 mg daily, and Sertraline 50 mg twice daily." The enhancement enables accurate interpretation of both standardized and colloquial medical terminology while maintaining clinical meaning, significantly improving transcription accuracy and reducing documentation errors in healthcare settings. By incorporating domain-specific medical knowledge during fine-tuning, the model effectively disambiguates and corrects ASR-generated errors, thereby improving the accuracy and reliability of automated medical transcription workflows. 

\begin{figure}[h]
    \centering
    \includegraphics[width=0.9\textwidth, height=0.25\textheight]{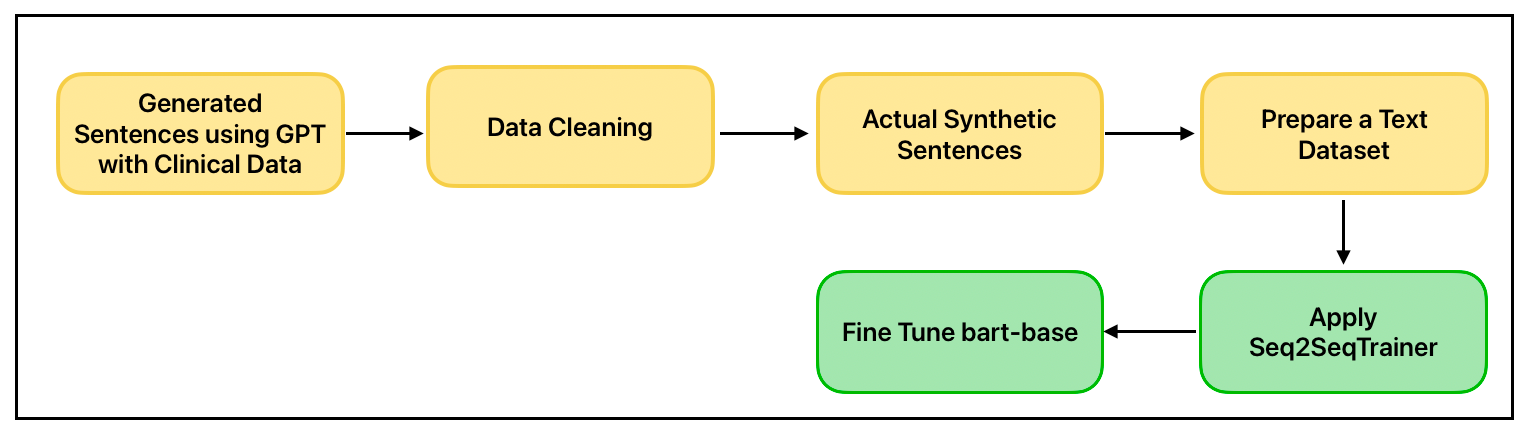}
    \caption{Synthetic Data Pipeline and United-MedASR Training Process
}
    \label{fig3}
\end{figure}

This model undergoes a separate but slightly similar training process as outlined for the Whisper model.
The Bart-base model is trained using the trainer with the cleaned and prepared medical text data. The training process involves defining appropriate parameters, such as learning rate 4\text{e-}4
 and training batch size 8, and applying them consistently throughout the training process as Figure \ref{fig3}.

After the training is completed, the model is evaluated on a validation or test dataset. The performance is measured using the Character Error Rate (CER), which provides a detailed assessment of the model's ability to correctly transcribe medical text. The Character Error Rate (CER) is a metric used to evaluate the performance of a system by comparing the recognised text with the reference text. It is calculated as the sum of the number of insertions, deletions, and substitutions needed to transform the recognised text into the reference text, divided by the total number of characters in the reference text.
\begin{equation}
\text{CER} = \frac{S + D + I}{N}
\end{equation}

\noindent Where:

\begin{itemize}
    \item $S$ = Number of substitutions (characters incorrectly recognised as something else)
    \item $D$ = Number of deletions (characters present in the reference text but missing in the recognised text)
    \item $I$ = Number of insertions (extra characters in the recognised text that are not in the reference text)
    \item $N$ = Total number of characters in the reference text
\end{itemize}

\noindent The CER is expressed as a percentage or a fraction, depending on how it is reported. A lower CER indicates better performance of the system.
The predictions made by the semantic enhancement model, along with its configurations, are saved for future reference. These predictions can further enhance the ASR system's accuracy, particularly in correcting semantic errors in medical transcriptions.
\begin{algorithm}[H]
\caption{Fine-Tune Bart-base Model Function}
\SetKwFunction{FFineTuneSpellingModel}{FineTuneSpellingModel}
\KwIn{$D_{train}, D_{test}$: Training and testing data}
\KwOut{$M_{semantic}, \text{CER}$: Fine-tuned semantic enhancement model and CER}
$M_{BART} \gets \textsc{LoadBARTBase}()$\;
$M_{semantic} \gets \textsc{FineTune}(M_{BART}, D_{train})$\;
$\text{CER} \gets \textsc{EvaluateCER}(M_{semantic}, D_{test})$\;
\Return $M_{semantic}, \text{CER}$\;
\end{algorithm}
\section{Evaluation}
\subsection{Synthetic Audio Data Quality Analysis}
The examination of the audio data, in settings unveils important features outlined in Table \ref{tab:audio_parameters}. The Signal-to-noise Ratio (SNR) measured at 0 dB signifies a balance between signal and noise levels. This implies that the audio file includes an amount of background noise compared to the signal. This might be deliberate to evaluate the effectiveness of speech recognition systems, in challenging conditions. An SNR, like that, is not ideal for real-world use since it could disrupt proper speech understanding in settings where precision matters most. A 24kHz sample rate works great for capturing all the nuances in terms without compromising on quality or processing speed. Moreover, a 384kbps bitrate ensures notch quality crucial, for maintaining the authenticity of synthesised speech in analysis. The combined factors highlight how well the audio is suited for testing and validating speech recognition models in situations that closely resemble world settings. 

\begin{table}[h]
\centering
\caption{Audio Parameters and Their Descriptions}
\label{tab:audio_parameters}
\begin{tabular}{|l|l|l|}
\hline
\textbf{Parameter} & \textbf{Value}       & \textbf{Description}                                          \\ \hline
\textbf{SNR (dB)}  & 0.00 dB              & Signal-to-Noise Ratio, indicating the audio clarity.          \\ \hline
\textbf{Sample Rate} & 24,000 Hz          & The number of samples of audio carried per second.            \\ \hline
\textbf{Bitrate}   & 384.00 kbps          & The amount of data processed per second in kilobits.          \\ \hline
\end{tabular}
\end{table}

\subsection{Training Evaluation}
\begin{figure}[h]
    \centering
    \includegraphics[width=0.6\textwidth, height=0.3\textheight]{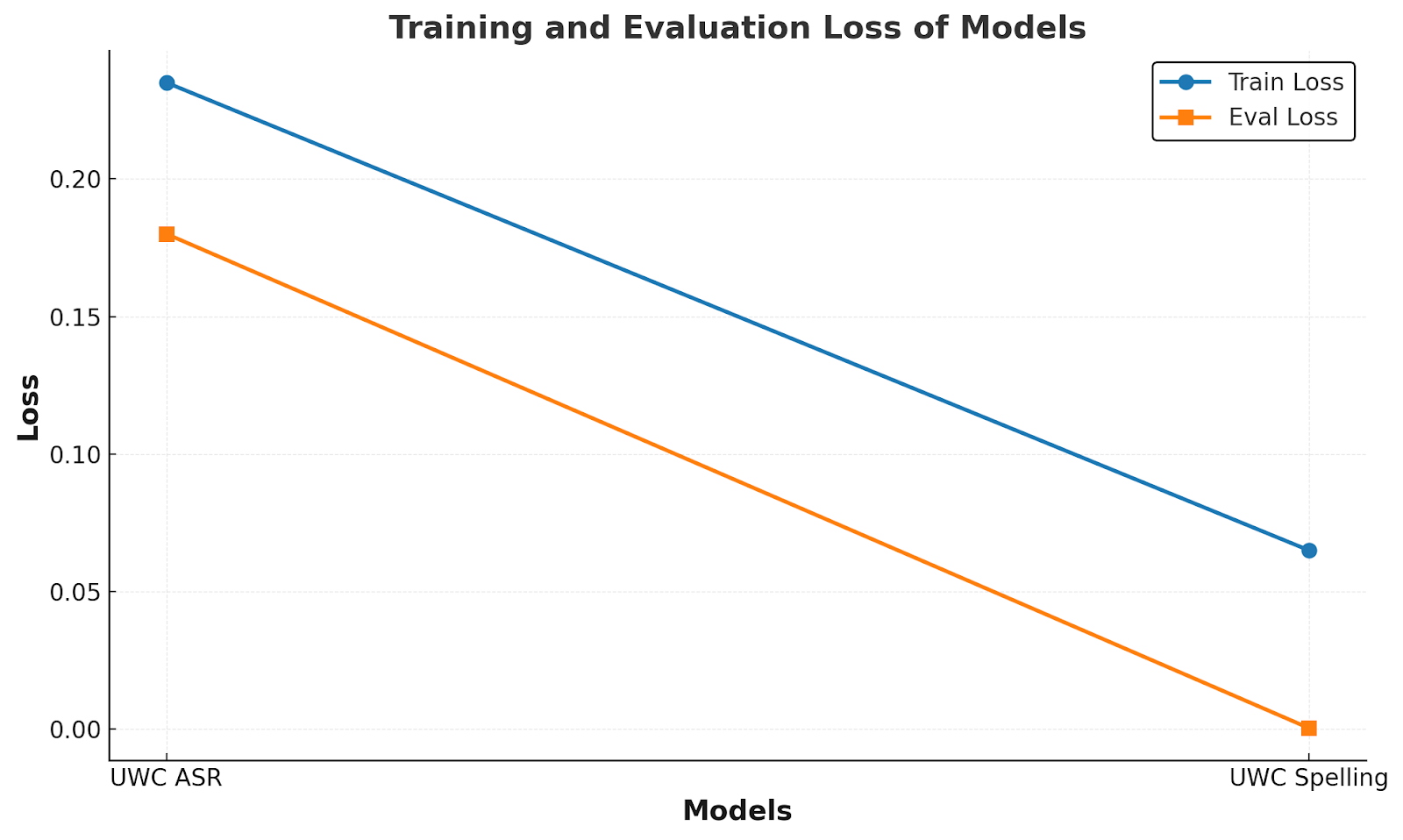}
    \caption{Performance Metrics of Fine Tuning of the Whisper and Bart-Base on Clinical Data.}
    \label{fig:fig4}
\end{figure}

Assessing how well a model performs involves determining the error rate, in both the training and evaluation phases of its development process. During training sessions at each stage of the process, a comparison of the model's predictions with outcomes helps in calculating loss which indicates how effectively the model is learning. As illustrated in Figure \ref{fig:fig4} presented earlier evaluation loss is derived from a dataset to measure how effectively the model can adapt to new scenarios. A low evaluation loss signifies performance when dealing with data whereas a high loss might indicate issues, like overfitting or underfitting. Both measurements play a role, in evaluating the model's performance and resilience. 

\subsection{ASR Benchmark Evaluation}
An Automatic Speech Recognition (ASR) benchmark is a standardised dataset used to evaluate the performance of ASR models. These benchmarks consist of audio recordings and their corresponding text transcriptions, serving as a reference for comparing the accuracy of speech-to-text systems. The LibriSpeech test-clean set, derived from audiobooks, is a widely used benchmark for ASR systems in English, providing a clean and high-quality speech dataset. The Europarl-ASR EN Guest-test set focuses on English transcriptions from European Parliament proceedings, challenging ASR models with real-world conversational speech. The Tedlium benchmark is based on TED talks, providing a rich variety of speech styles, accents, and topics, making it a valuable resource for assessing ASR systems' robustness to different speaking conditions. Together, these benchmarks cover a broad range of speech scenarios, allowing for a comprehensive evaluation of ASR performance. Lastly, the FLEURS dataset, designed for multilingual speech recognition, includes over 100 languages, testing ASR models' ability to generalize across diverse linguistic contexts. These datasets enable researchers to compute important metrics such as the Word Error Rate (WER), which reflects the accuracy of the ASR system by measuring the differences between the generated and actual transcriptions. We applied a series of text transformations to both the actual and transcribed text. First, all the text was converted to lowercase to ensure consistency. Then, we removed any extra spaces and replaced them with a single space to standardise spacing. Following that, any sequences of multiple spaces were reduced to just one. Finally, the text was split into lists of words based on spaces, making it easier to compare the actual and transcribed versions. Benchmarks like these are essential for improving the robustness and performance of ASR systems across various speech domains and noise conditions.
\begin{algorithm}[H]
\caption{Evaluate on Benchmarks Function}
\SetKwFunction{FEvaluateOnBenchmarks}{EvaluateOnBenchmarks}
\KwIn{$M_{UW}, M_{semantic}$: ASR model and Semantic Enhancement model }
\KwOut{$R$: Results from benchmarks}
$B \gets \{\text{LibriSpeech}, \text{Europarl-ASR}, \text{TED-LIUM}, \text{FLEURS}\}$\;
$R \gets \emptyset$\;
\For{$b \in B$}{
    $D_b \gets \textsc{LoadBenchmark}(b)$\;
    $D_{b,clean} \gets \textsc{ApplyNoiseReduction}(D_b)$\;
    $T_b \gets M_{UW}(D_{b,clean})$\;
    $T_{b,corr} \gets M_{semantic}(T_b)$\;
    $\text{WER}_b \gets \textsc{CalculateWER}(T_{b,corr}, D_b)$\;
    $R \gets R \cup \{(b, \text{WER}_b)\}$\;
}
\Return $R$\;
\end{algorithm}
\subsubsection{The Process of  Benchmark Evaluation}
\begin{figure}[h]
    \centering
    \includegraphics[width=0.8\textwidth, height=0.2\textheight]{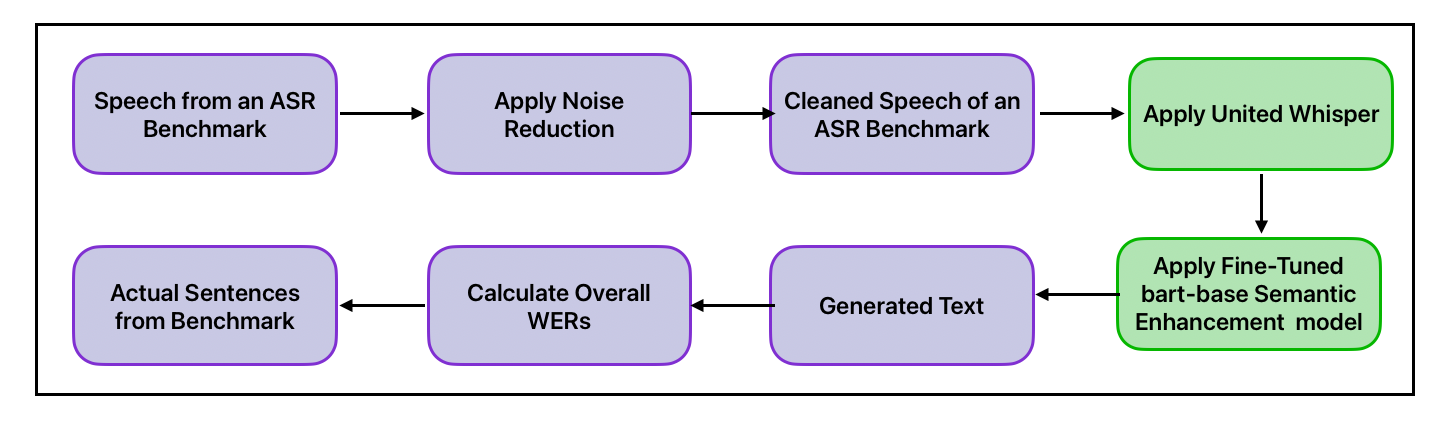}
    \caption{ United-MedASR Benchmarks Evaluation Flow.}
    \label{fig:fig5}
\end{figure}

We assessed the LibriSpeech test dataset alongside Europarl ASR-EN Guest test data, FLEURS and Tedlium datasets using the process outlined in Figure \ref{fig:fig5}. 
This figure depicted demonstrates the workflow that starts with feeding speech data from well-known datasets and then applying an algorithm for reducing noise interference, in the input data stream \cite{37}\cite{38}. The information is subsequently fed into the United-MedASR model, for the transcription process. Ultimately, the Word Error Rate (WER) is calculated by comparing the transcribed text with reference sentences, providing an accurate measure of the model’s performance and transcription accuracy across different datasets. After evaluating the Word Error Rate, for each sample and then averaging them out to assess the model's overall performance effectively gauges its capability to manage a variety of spoken inputs. 

\begin{figure}[h]
    \centering
    \includegraphics[width=0.9\textwidth, height=0.4\textheight]{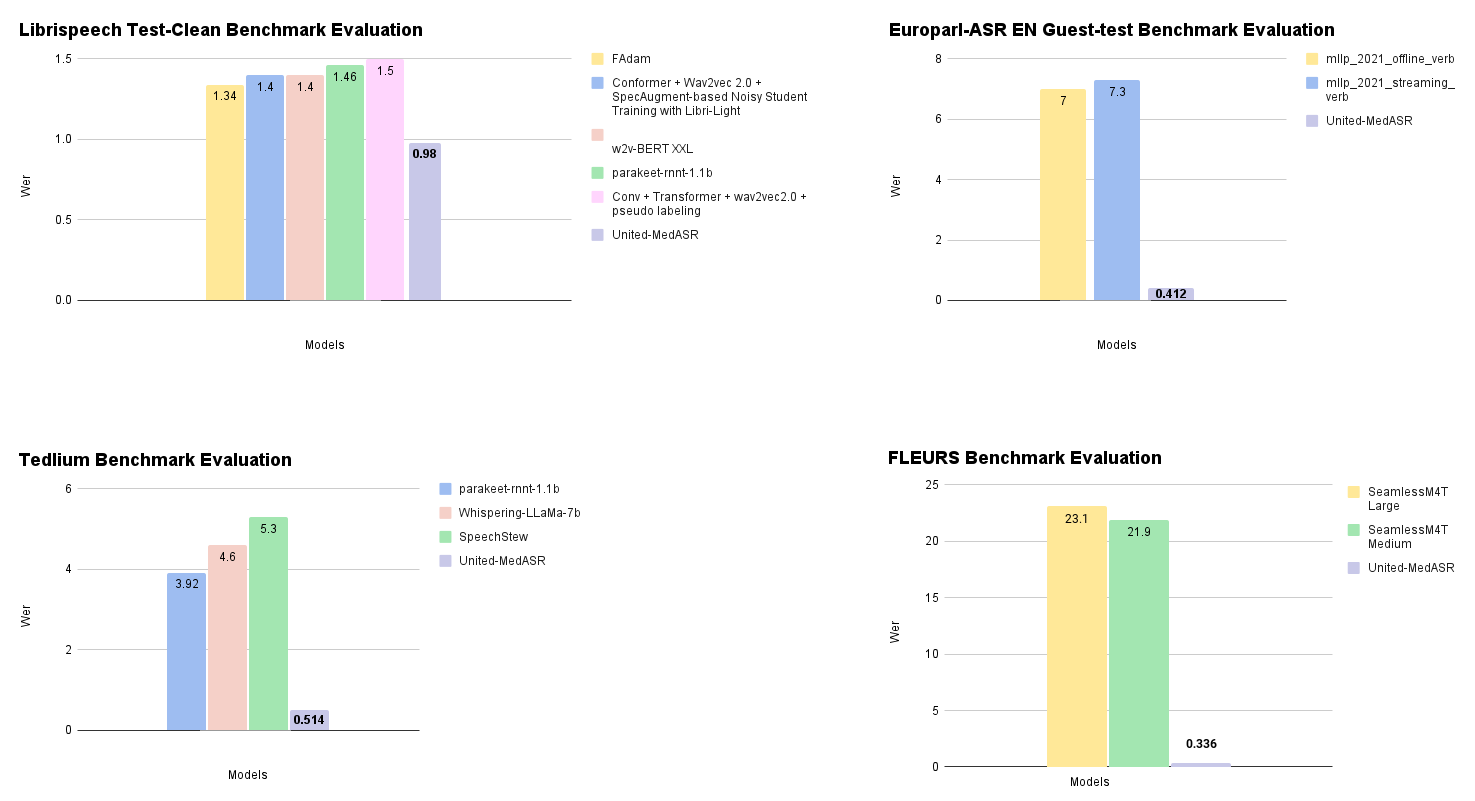}
    \caption{The Word Error Rate (WER) performance of different models on four benchmark datasets: (a) Librispeech Test-Clean, (b) Europarl-ASR EN Guest-test, (c) Tedlium, and (d) FLEURS.}
    \label{fig:fig6}
\end{figure}

Figure \ref{fig:fig6} and Table \ref{tab:asr_model_performance} shows the assessment results, for the United-MedASR model using ASR benchmarks. The outcomes emphasise how various models perform across benchmarks by comparing their Word Error Rate (WER) scores as the accuracy measure. In the graphs, bars are model performances in terms of WER, in each corresponding benchmark. 

\subsubsection{LibriSpeech test-clean Benchmark Evaluation}
In our evaluation of the LibriSpeech test-clean benchmark \cite{33}, the United-MedASR achieved an impressive Word Error Rate (WER) of 0.985, demonstrating exceptional transcription accuracy and outperforming other models on the same dataset. The LibriSpeech test-clean, a key ASR benchmark, provides a rigorous test environment with diverse, high-quality speech data.  The low WER reflects the models' ability to handle real-world speech variations, including pronunciation, accents, and background noise, affirming the effectiveness of combining ASR techniques with semantic enhancement. Table \label{tab:asr_model_performance} compares these results, establishing the United-MedASR model as a leading ASR solution and setting a new standard for future developments.


\subsubsection{Europarl-ASR EN Guest-test Benchmark Evaluation}
In our evaluation of the Europarl-ASR EN Guest-test dataset using the United-MedASR model, we achieved an outstanding Word Error Rate (WER) of 0.412, significantly outperforming the previous method's WER of 7.00. Table \ref{tab:asr_model_performance} presents the performance of various ASR models over time for this benchmark \cite{35}.

The Europarl-ASR dataset, containing over 1,300 hours of parliamentary debates and 70 million tokens, includes both automatically noise-filtered and verbatim transcripts to enhance training. With 18 hours of manually verbatim transcripts for reliable speaker-dependent and independent test sets, it is a key resource for benchmarking ASR systems.

Our model's lower WER reflects its robustness in handling complex speech patterns, demonstrating the effectiveness of our noise-filtering and verbatim techniques. This result affirms the Europarl-ASR corpus as a vital benchmark for real-world ASR advancements.

\subsubsection{TED-LIUM 3 Benchmark Evaluation}

In the TEDLIUM benchmark evaluation \cite{48}, several speech recognition models from different years were compared, including our model, United-MedASR, evaluated in 2024. The 2023 models, such as parakeet-rnnt-1.1b and Whispering-LLaMa-7b, achieved Word Error Rates (WER) of 3.92 and 4.60, respectively, while the 2021 model SpeechStew had a WER of 5.30. These models demonstrated varying levels of accuracy in speech-to-text transcription. In contrast, our United-MedASR model, evaluated in 2024, achieved a notably lower WER of 0.514, indicating a significant improvement in accuracy compared to the earlier models. This result highlights the advancements made by our model in minimising transcription errors, positioning it as a more precise solution in the field of automatic speech recognition.

\subsubsection{FLEURS (English) Benchmark Evaluation}
In the FLEURS benchmark evaluation \cite{36}, two models from 2023, SeamlessM4T Large and SeamlessM4T Medium, were assessed alongside our model, United-MedASR, in 2024. The SeamlessM4T Large and Medium models achieved Word Error Rates (WER) of 23.1\% and 21.9\%, respectively, reflecting their accuracy in speech-to-text tasks during the 2023 evaluation. In contrast, our model, United-MedASR, demonstrated a significant improvement in performance with a WER of just 0.336\% for English, representing a major advancement in reducing transcription errors as Table \label{tab:asr_model_performance}. This result underscores the enhanced capabilities of our model compared to previous state-of-the-art approaches, indicating its potential for delivering more accurate speech recognition solutions.

\begin{table}[h]
\centering
\caption{Comparison of ASR Model Performance of all the Benchmarks}
\label{tab:asr_model_performance}
\resizebox{\textwidth}{!}{ 
\begin{tabular}{|p{2.5cm}|p{3.0cm}|p{1.7cm}|p{1.2cm}|p{1.2cm}|p{6cm}|p{1.5cm}|}
\hline
\textbf{Benchmarks} & \textbf{Domain} & \textbf{Multilingual} & \textbf{Multi-Speaker} & \textbf{Hours} & \textbf{Models} & \textbf{WER} \\ \hline

\multirow{6}{*}{LibriSpeech Test} & \multirow{6}{*}{Audiobook} & \multirow{6}{*}{No} & \multirow{6}{*}{Yes} & \multirow{6}{*}{5.4} & \raggedright FAdam\cite{16} & 1.34 \\ \cline{6-7} 
 &  &  &  &  & \raggedright Conformer + Wav2vec 2.0 + SpecAugment-based Noisy Student Training with Libri-Light\cite{43} & 1.4 \\ \cline{6-7} 
 &  &  &  &  & \raggedright w2v-BERT XXL\cite{42} & 1.4 \\ \cline{6-7} 
 &  &  &  &  & \raggedright parakeet-rnnt-1.1b\cite{17} & 1.46 \\ \cline{6-7} 
 &  &  &  &  & \raggedright Conv + Transformer + wav2vec2.0 + pseudo labeling\cite{44} & 1.5 \\ \cline{6-7} 
 &  &  &  &  & \textbf{United-MedASR} & \textbf{0.98} \\ \hline

\multirow{3}{*}{Europarl-ASR} & \multirow{3}{*}{Parliamentary sessions} & \multirow{3}{*}{No} & \multirow{3}{*}{Yes} & \multirow{3}{*}{3.9} & \raggedright mllp\_2021\_offline\_verb\cite{35} & 7 \\ \cline{6-7} 
 &  &  &  &  & \raggedright mllp\_2021\_streaming\_verb\cite{35} & 7.3 \\ \cline{6-7} 
 &  &  &  &  & \textbf{United-MedASR} & \textbf{0.412} \\ \hline

\multirow{4}{*}{TED-LIUM} & \multirow{4}{*}{TED talks} & \multirow{4}{*}{No} & \multirow{4}{*}{Yes} & \multirow{4}{*}{\textasciitilde452} & \raggedright parakeet-rnnt-1.1b\cite{17} & 3.92 \\ \cline{6-7} 
 &  &  &  &  & \raggedright Whispering-LLaMa-7b\cite{45} & 4.6 \\ \cline{6-7} 
 &  &  &  &  & \raggedright SpeechStew\cite{46} & 5.3 \\ \cline{6-7} 
 &  &  &  &  & \textbf{United-MedASR} & \textbf{0.514} \\ \hline

\multirow{3}{*}{Fleurs} & \multirow{3}{*}{Read speech} & \multirow{3}{*}{Yes} & \multirow{3}{*}{Yes} & \multirow{3}{*}{\textasciitilde19} & \raggedright SeamlessM4T Large\cite{47} & 23.1 \\ \cline{6-7} 
 &  &  &  &  & \raggedright SeamlessM4T Medium\cite{47} & 21.9 \\ \cline{6-7} 
 &  &  &  &  & \textbf{United-MedASR} & \textbf{0.336 (English)} \\ \hline

\end{tabular}
}
\end{table}

\subsubsection{Error Rates (WER) Distribution of  Benchmarks}
Table \ref{tab:wer_comparison} the comparison of Word Error Rates (WER) between four evaluated datasets - Europarl-ASR EN Guest-test, Librispeech Test-Clean, TEDLIUM, and FLEURS. The WER provides a critical measure of accuracy in automatic speech recognition (ASR) systems, where a lower WER indicates higher transcription accuracy. 
The table shows that the Europarl-ASR dataset exhibits a lower and more consistent WER distribution, with values closer to the lower end of the range (between approximately 0.1 and 0.75). This indicates a relatively higher performance in transcription accuracy compared to Librispeech, which shows a wider spread in WER values (ranging from 0.3 to 1.3), with a median that is higher than Europarl's. The TEDLIUM dataset exhibits a broad range of WER values, from 0 to 11, with a median of 0.51, indicating high variability in transcription performance. This spread suggests that ASR systems struggle with TEDLIUM due to factors like diverse speech patterns and potential background noise in TED talks. Some parts of the dataset are transcribed well, but others pose significant challenges.

In contrast, the FLEURS dataset has a narrower WER range (0 to 1.2) and a lower median of 0.34, implying better and more consistent transcription accuracy. The tighter distribution reflects the dataset's clearer speech and higher recording quality, making it easier for ASR systems to handle.



\begin{table}[h]
\centering
\caption{Comparison of Minimum, Maximum, and Median Word Error Rates (WER) Across Four Evaluated Benchmarks in Automatic Speech Recognition (ASR) Systems}
\label{tab:wer_comparison}
\begin{tabular}{|l|c|c|c|}
\hline
\textbf{Benchmarks} & \textbf{WER Minimum} & \textbf{WER Maximum} & \textbf{Median WER} \\ \hline
Europarl-ASR EN Guest-test & 0.1 & 0.75 & 0.26 \\ \hline
Librispeech Test-Clean     & 0.3 & 1.3  & 0.98 \\ \hline
TEDLIUM                    & 0   & 11   & 0.51 \\ \hline
FLEURS                     & 0   & 1.2  & 0.34 \\ \hline
\end{tabular}
\end{table}

\subsection{Comparison with General-Purpose and Medical-Specific ASR Models}
In comparison to state-of-the-art medical ASR systems such as Nuance's Dragon Medical One, MModal Fluency, and general-purpose ASR models like Google Speech-to-Text, Amazon Transcribe, and Microsoft Azure Speech-to-Text, United-MedASR demonstrates significant architectural advantages due to its transformer-based model \cite{6},\cite{7},\cite{39}-\cite{41}. While Dragon Medical One and MModal Fluency are designed specifically for medical transcription tasks, and general-purpose models such as Google Speech-to-Text, Amazon Transcribe, and Microsoft Azure Speech-to-Text excel in a wide range of applications, they fall short in handling domain-specific medical terminology. These general models are not fine-tuned for medical transcription, often resulting in high Word Error Rates (WER) when tasked with complex medical jargon. United-MedASR, on the other hand, leverages the self-attention mechanism of transformers, enabling it to capture long-range dependencies and context crucial for transcribing specialised medical vocabulary accurately. The fine-tuning of United-MedASR on synthetic medical speech datasets, including those generated from ICD-10 and FDA, further improves its ability to adapt to medical-specific terms, setting it apart from these general-purpose ASR models, which typically require manual updates or additional configuration to accommodate new or specialised terminology. In addition, United-MedASR integrates a fine-tuned BART-base model for semantic enhancement, improving transcription accuracy for complex medical terms and ensuring superior performance over general ASR models in healthcare settings. Its superior performance across several benchmarks reflects these advantages, achieving a 0.985 WER on the LibriSpeech test-clean dataset, making it a more reliable choice for medical transcription tasks compared to general-purpose ASR systems.

\section{Discussion}
United-MedASR has shown performance, in datasets and is now being used in actual clinical scenarios like at United We Care's clinical assistance programs. However, there are some aspects to consider for its wider use. In real-life tests, the results have been positive. More examination is necessary to gauge how well the system works in changing and noisy medical settings. The use of data for training purposes brings up worries about privacy and the risk of de-anonymization particularly in sensitive fields, like healthcare. Complying with regulations such, as HIPAA and GDPR is vital to safeguarding patient privacy confidentiality is essential in the healthcare sector. Moreover presenting details in a more straightforward manner can help a wider audience understand the study outcomes better. Especially healthcare professionals when discussing topics like transformer architectures and Word Error Rate (WER). Lastly, we must consider the ongoing viability of using data, in the field given how medical terminology constantly evolves. Consistently improving the model to adjust to evolving terminologies and controlling expenses is crucial, for guaranteeing the scalability and long-term significance of United-MedASR, in environments. 

\section{Future Works}
Building upon United-MedASR's success in medical ASR, several promising research directions emerge for enhancing the system's adaptability and robustness. A primary avenue for advancement lies in extending the current BART-base semantic enhancer to incorporate zero-shot learning capabilities, enabling real-time adaptation to new medical terminology. This enhancement would involve developing a context-aware pattern recognition system that identifies valid new medical terms based on morphological features and usage patterns, complemented by a confidence scoring mechanism that evaluates new terms through multiple linguistic dimensions such as context, morphology, frequency, and source reliability. An automated validation system could cross-reference newly encountered terms with emerging medical literature and clinical databases, while a dynamic vocabulary update mechanism would maintain system accuracy while incorporating verified new terms.

To improve United MedASR's performance in handling terms and uncommon illnesses effectively we propose exploring innovative methods to enable zero-shot ASR capabilities. This includes implementing a phoneme-based recognition system that can decompose and reconstruct new medical terms from known phonemic patterns, and integrating cross-modal learning that leverages text resources such as medical journals and clinical notes to inform acoustic model adaptation. A context-aware inference system could use surrounding clinical terminology to enhance recognition of unknown terms, while a specialised medical phoneme inventory would capture unique pronunciation patterns in clinical settings.

\bibliographystyle{plain}  


\end{document}